\documentclass{elsart1p}
\usepackage{graphicx}
\usepackage{amssymb}
\usepackage{bbm}
\usepackage{amsmath}
\usepackage{amsfonts,amsbsy}

\newcommand{\dsigmap}{{\frac{\ud \sigma^\textrm{p}_\textrm{dip}}{\ud^2 \bt}}}
\newcommand{\qscgc}{{\bar Q}_\mathrm{s}}
\newcommand{\ud}{\, \mathrm{d}}

\newcommand{\qssf}{{\tilde{Q}^2_\mathrm{s,p}}}

\newcommand{\rt}{{\mathbf{r}_\perp}}
\newcommand{\kt}{{\mathbf{k}_\perp}}
\newcommand{\pt}{{\mathbf{p}_\perp}}

\newcommand{\bt}{{\mathbf{b}_\perp}}

\newcommand{\nc}{{N_\mathrm{c}}}

\begin{document}
\begin{frontmatter}
%
%
%
%
%
\title{Inclusive hadron distributions in p+p collisions from saturation models of HERA DIS data}
%
\author[a1]{Prithwish Tribedy}
\author{and}
\author[a2]{Raju Venugopalan}
 \address[a1]{Variable Energy Cyclotron Centre, 1/AF Bidhan Nagar, Kolkata-700064, India}
 \address[a2]{Physics Dept., Bldg. 510A, Brookhaven National Laboratory, Upton, NY 11973, USA}
%

\begin{abstract}
Dipole models based on various saturation scenarios provide reasonable fits to small-x DIS inclusive, diffractive and exclusive data from HERA. Proton un-integrated gluon distributions extracted from such fits are employed in a $k_\bot$-factorization framework to calculate inclusive gluon distributions at various energies. The n-particle multiplicity distribution predicted in the Glasma flux tube approach shows good agreement with data over a wide range of energies. Hadron inclusive transverse momentum distributions expressed in terms of the saturation scale demonstrate universal behavior over a wider kinematic range systematically with increasing center of mass energies. 
\end{abstract}

\end{frontmatter}

\section{Introduction}
HERA deeply inelastic scattering (DIS) results on structure functions demonstrate a rapid  bremsstrahlung growth of the gluon density at small x. When interpreted in the same framework as the parton model, this growth is predicted to saturate because the gluon occupation number in hadron wave functions saturate at a value maximally of order $1/\alpha_S$; dynamically, nonlinear effects such as gluon recombination and screening by other gluons deplete the growth of the gluon distribution\cite{GLR}.  Gluon modes with $k_T < Q_S(\gg \Lambda_{QCD})$ are maximally occupied, where $Q_S^2(x)$ is a dynamically generated semi-hard scale called the saturation scale. For small $x$, $Q_S(x)$ is large enough that high occupancy states can be described by weak coupling classical effective theory\cite{MV}. This Color Glass Condensate description of high energy hadrons and nuclei is universal and has been tested in both DIS and hadronic collisions. In particular, saturation based phenomenological predictions successfully describe recent LHC p+p data \cite{Lev-Rez,PTRV} and predict possible geometrical scaling of transverse momentum distribution\cite{Larry-Michal,Praszalowicz} similar to the geometrical scaling observed previously in DIS.

The object common to DIS and hadronic collisions is the dipole cross section $\dsigmap\left(\rt,z,\bt\right)$.  In the CGC framework, the dipole cross section can be expressed in terms of expectation values of correlators of Wilson lines representing the color fields of the target. The energy dependence of this quantity comes from renormalization group evolution but to get the realistic impact parameter dependence one has to rely on models involving parametrizations constrained by experimental data. In the large $\nc$ limit, the dipole cross section is related to the un-integrated gluon distribution inside hadron/nucleus as

\begin{equation}
\frac{\textmd{d}\phi(x,\textbf{k}_{\bot}|\textbf{s}_{\bot})}{\textmd{d}^2\textbf{s}_{\bot}} =\frac{\textbf{k}_\bot^2 N_c}{4 \alpha_S}  \int \limits_{0}^{+\infty}\textmd{d}^2\textbf{r}_{\bot}
e^{i \textbf k_{\bot}.\textbf{r}_{\bot}} \left[1 - \frac{1}{2}\, \frac{\ud \sigma^\textrm{p}_\textrm{dip}}{\ud^2 \textbf{s}_\perp\\} (\rt,x,\textbf{s}_\perp)\right]^{2}.
\label{eq:unint-gluon}
\end{equation}
  
 For hadron-hadron collisions, the inclusive gluon distribution which is $\kt$-factorizable into the products of un-integrated gluon distributions in the target and projectile is expressed as
\begin{equation}
 \frac{\textmd{d}N_{g}(\textbf{b}_{\bot})}{\textmd{d}y~\textmd{d}^{2}\textbf{p}_{\bot}}=\frac{16 \alpha_S}{\pi C_F} \frac{1}{p_{\bot}^2} 
\int \frac{\textmd{d}^{2} \textbf{k}_{\bot}}{(2\pi)^{5}} \int \textmd{d}^{2} \textbf{s}_{\bot} \frac{\textmd{d}\phi_A(x_1,\textbf{k}_{\bot}|\textbf{s}_{\bot})}{\textmd{d}^2\textbf{s}_{\bot}}
\frac{\textmd{d}\phi_B(x_2,\textbf{p}_{\bot}-\textbf{k}_{\bot}|\textbf{s}_{\bot}-\textbf{b}_{\bot})}{\textmd{d}^2\textbf{s}_{\bot}}.
\label{eq:ktfact1}
\end{equation}

\section{Saturation models of HERA DIS}

Two models of the dipole cross-section that have been extensively compared to 
HERA data are the IP-Sat~\cite{KT,KMW} and the b-CGC~\cite{IIM,KW} models. In the former the impact parameter dependence is introduced through  a normalized Gaussian profile function $T_p(\bt)$ and in the latter through a scale $\qscgc (x,\bt)$. For a detailed discussion of the parameters involved in these models and their values from fits to HERA data, see ref.~\cite{PTRV}.
\begin{figure}[htl]
\centerline{
\includegraphics[height=5cm,width =6cm]{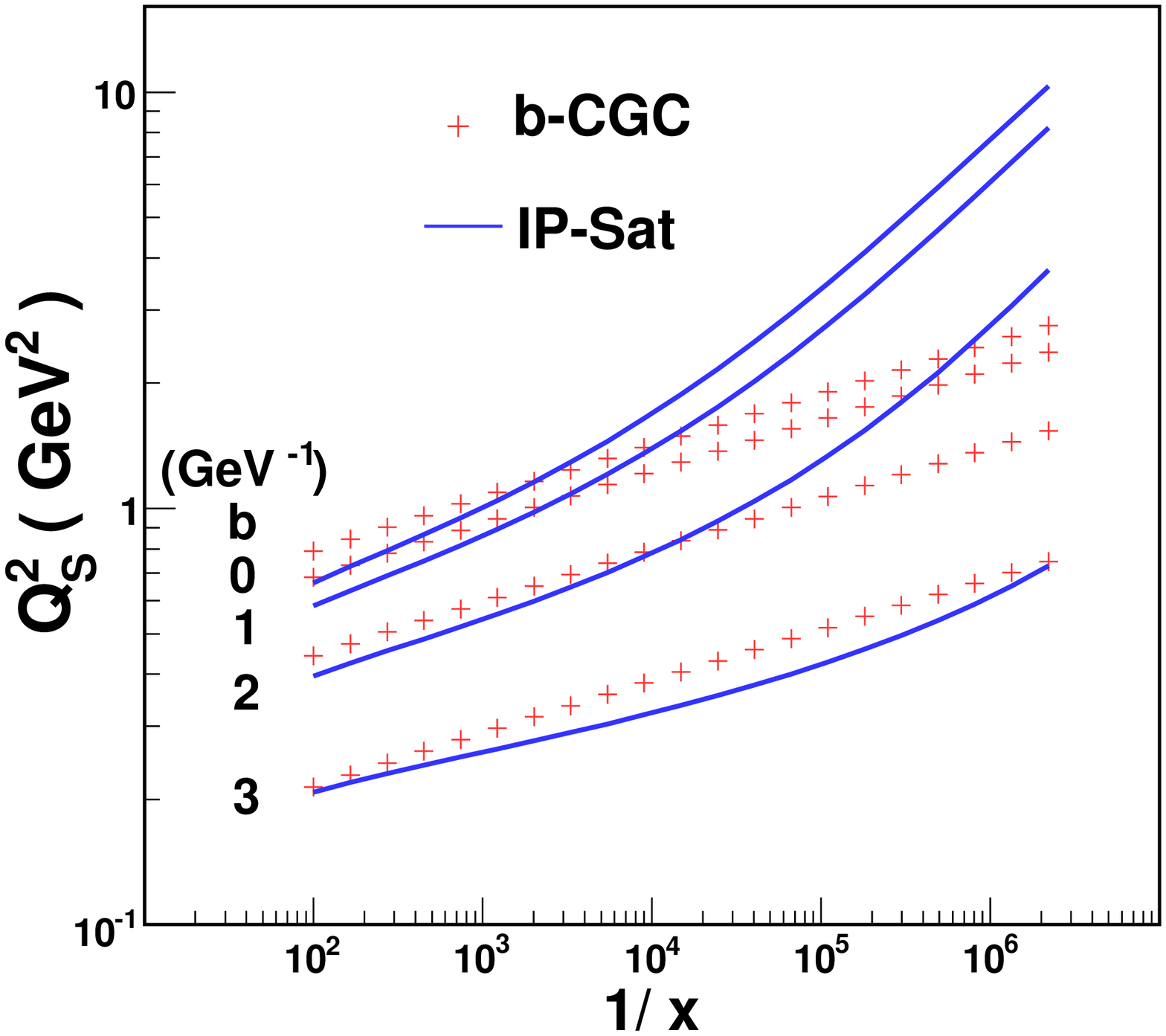}
\includegraphics[height=5cm,width =6cm]{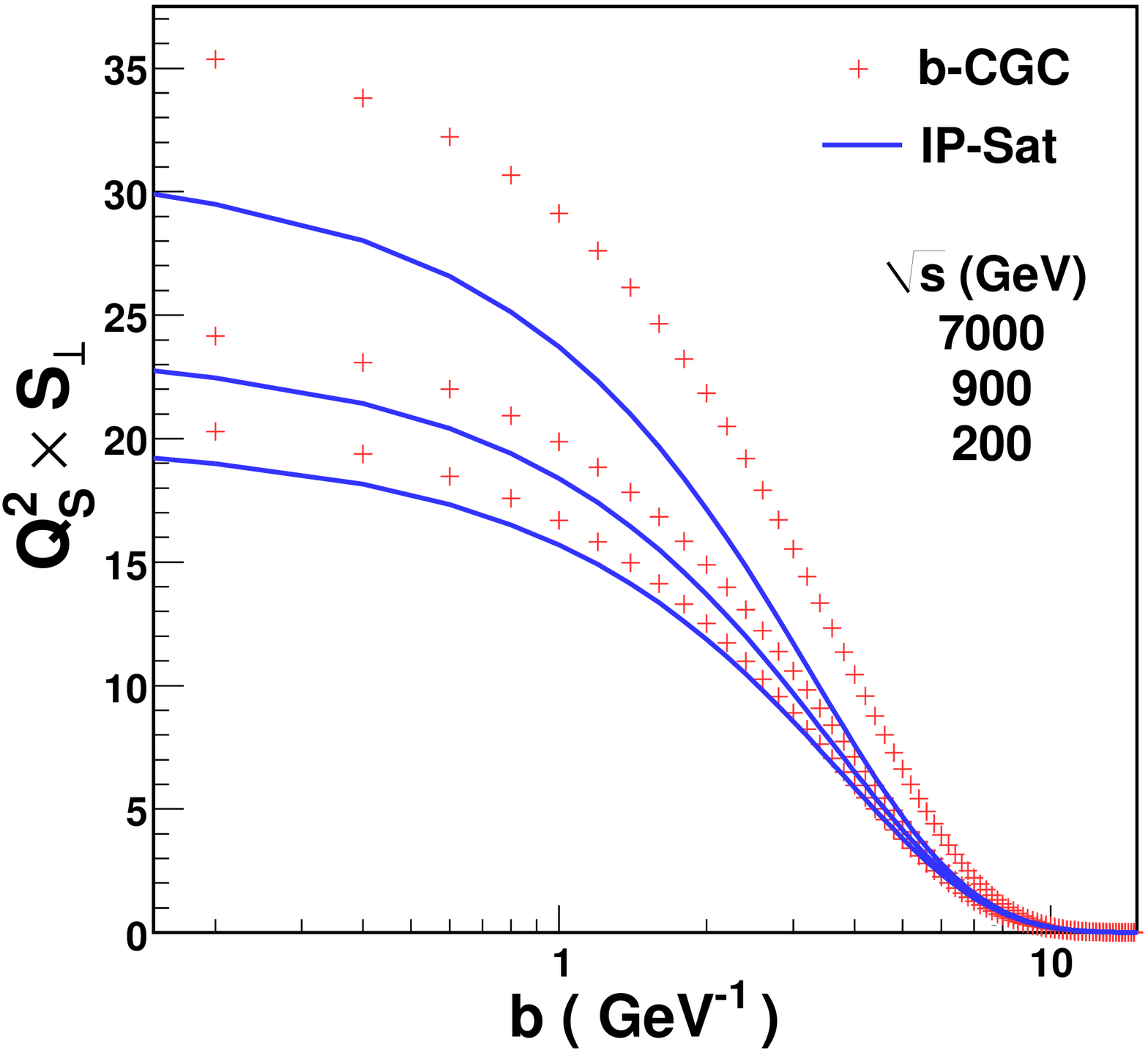}
}
\caption{Left: Adjoint saturation scales obtained from the IP-Sat(blue lines) and b-CGC models(red crosses). Right: Impact parameter and energy dependence of $Q_S^2 S_\perp$.}
\label{fig:satscale}
\end{figure}
The saturation scale in the fundamental representation for both the models can be calculated self consistently solving $\dsigmap[x,\rt^2=1/\qssf(x,\bt)]=2(1-e^{-1/2})$. The corresponding adjoint saturation scale $Q_S^2$, relevant for hadronic collisions, is obtained by multiplying $\qssf$ by 9/4. In the range $x\approx 10^{-2}$-$10^{-4}$, the behaviour of $Q_S^2$ (see fig.\ref{fig:satscale} left) at $b=0$ can be approximated by a function of the form $(x_0/x)^\lambda$ with $\lambda\sim 0.12$ for the b-CGC model and $\lambda\sim 0.2$ for the IP-Sat model.
\begin{figure}[htl]
\centerline{
\includegraphics[height=5cm,width =6cm]{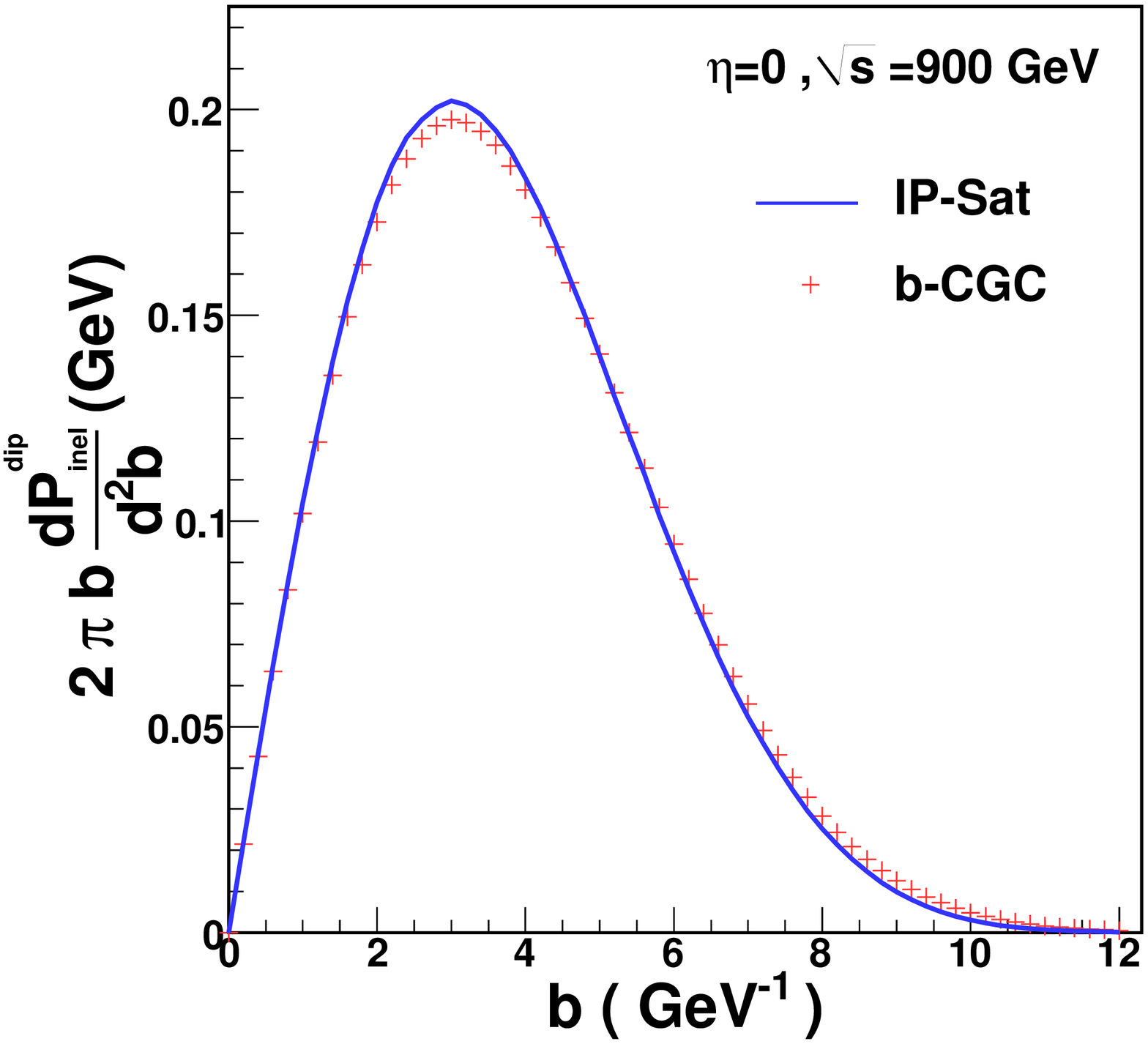}
\includegraphics[height=5cm,width =6cm]{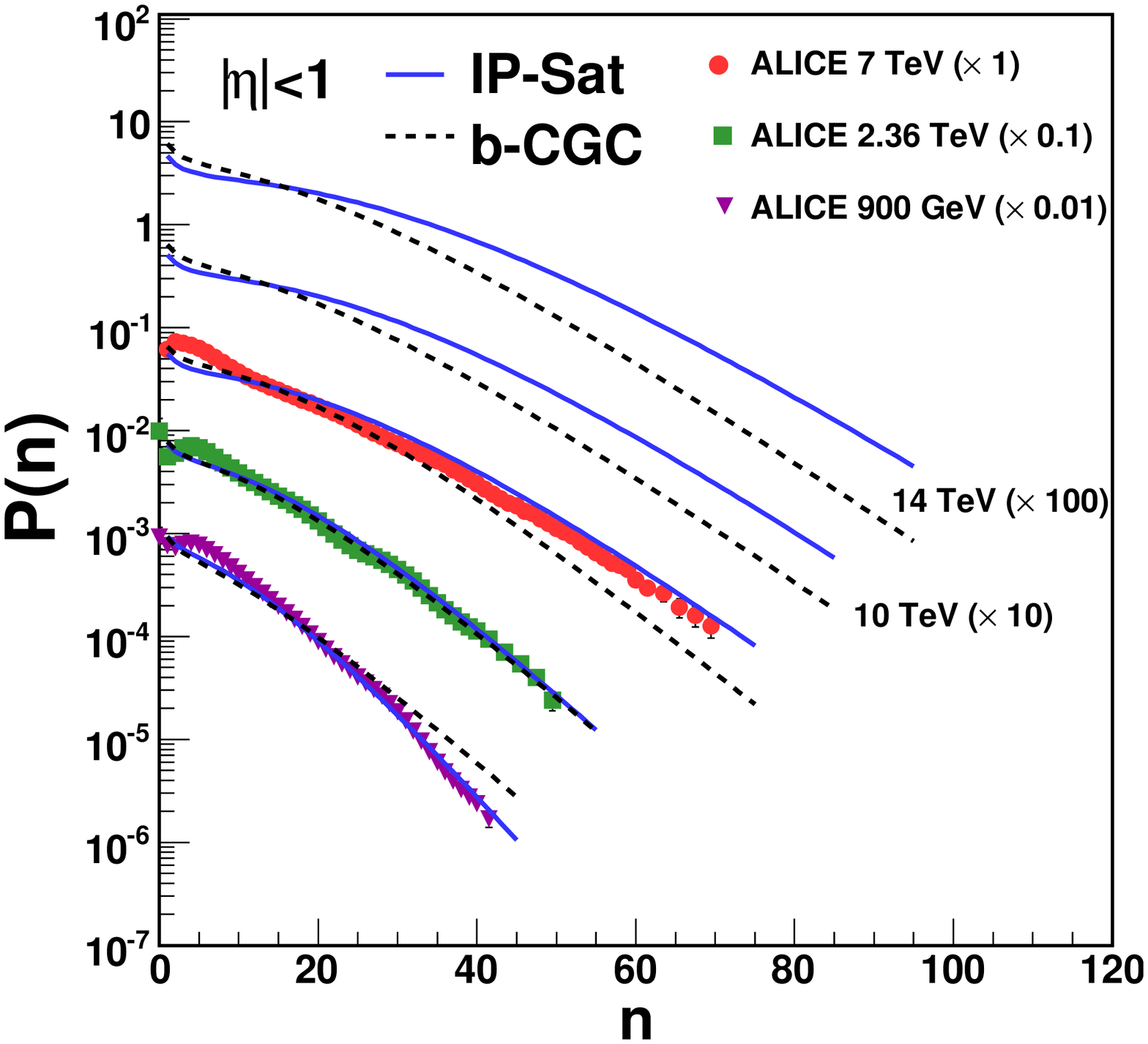}
}
\label{fig:multdist}
\caption{Left: Impact parameter dependence of the probability distribution for inelastic collision. Right: Charged-particle multiplicity distributions in $|\eta|< 1.0$ as predicted from saturation models, data points are from Ref\cite{ALICE-Aamodt}}
\end{figure}
\section{Results and discussion}
Multiparticle production in high energy hadronic collisions can be treated self consistently in the  CGC approach. The Glasma flux tube picture~\cite{DumitGMV} predicts~\cite{GelisLM}  that the n-particle correlation is generated by the negative binomial distribution $P_n^{\rm NB}({\bar n},k)$. It is characterized by two parameters, the mean multiplicity $\bar n$ and $k$. At a given impact parameter of the  collision, the mean multiplicity $\bar n \approx\bar n(b)$ is obtained by integrating eq.~\ref{eq:ktfact1} over $\pt$. In the Glasma picture, the parameter $k(b)=\zeta (N_c^2-1) Q_S^2 S_\perp/ 2\pi$ with $\zeta \sim {\cal O}(1)$ \cite{LappiSV}. The quantity $Q_S^2 S_\perp$ shown in fig.\ref{fig:satscale} (right) is the number of flux tubes in the overlap area $S_\perp$ of two hadrons. Convolving $P_n^{\rm NB}({\bar n(b)},k(b))$ with the probability distribution ${dP_{\rm inel.} \over d^2\bt}$ for an inelastic collision at $b$-fig.~\ref{fig:multdist} (left)-one obtains \cite{PTRV} the n-particle inclusive multiplicity distribution as shown in fig.~\ref{fig:multdist} (right). 

\begin{figure}[t]
\centerline{
\includegraphics[height=5cm,width =6cm]{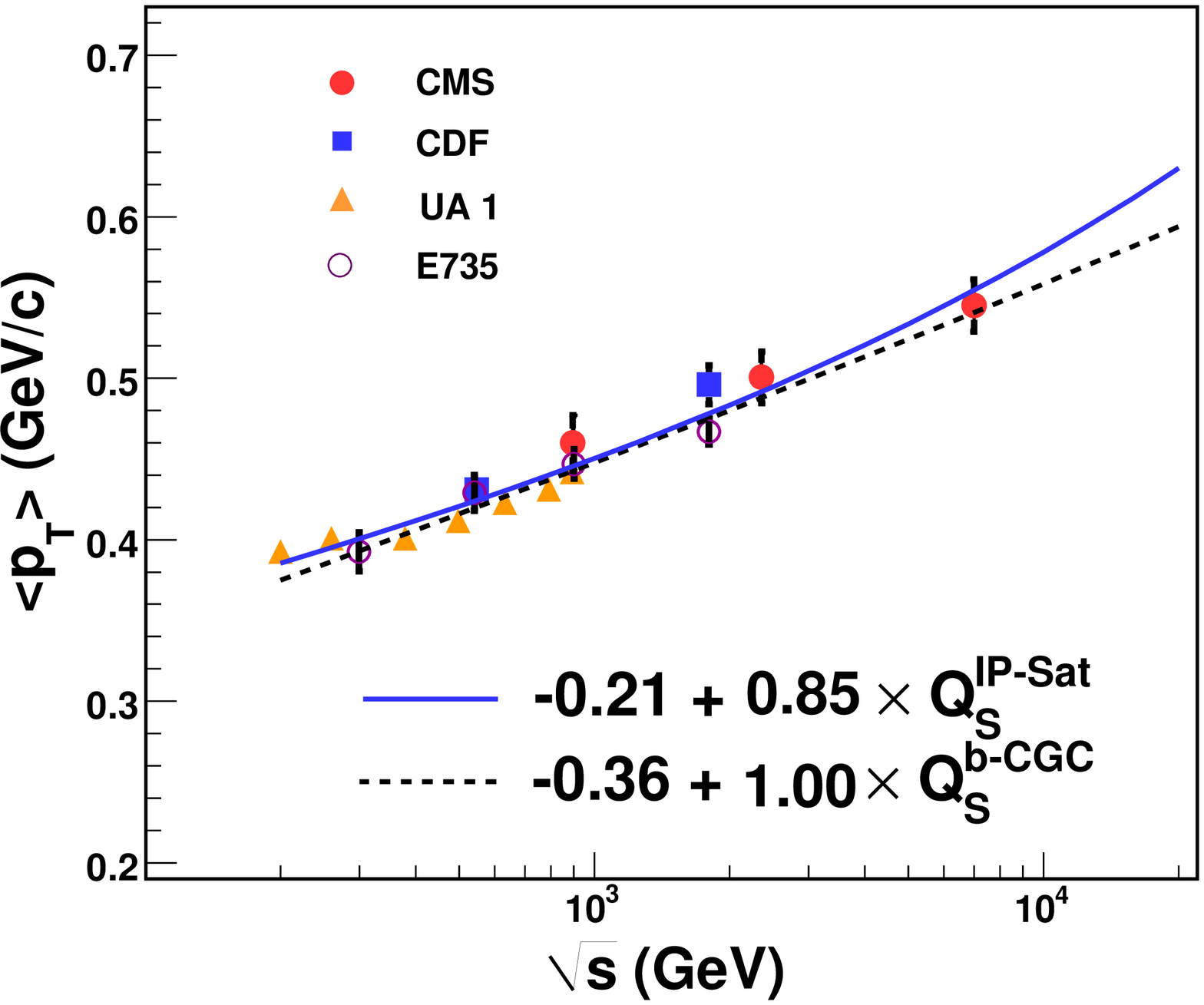}
\includegraphics[height=5cm,width =6cm]{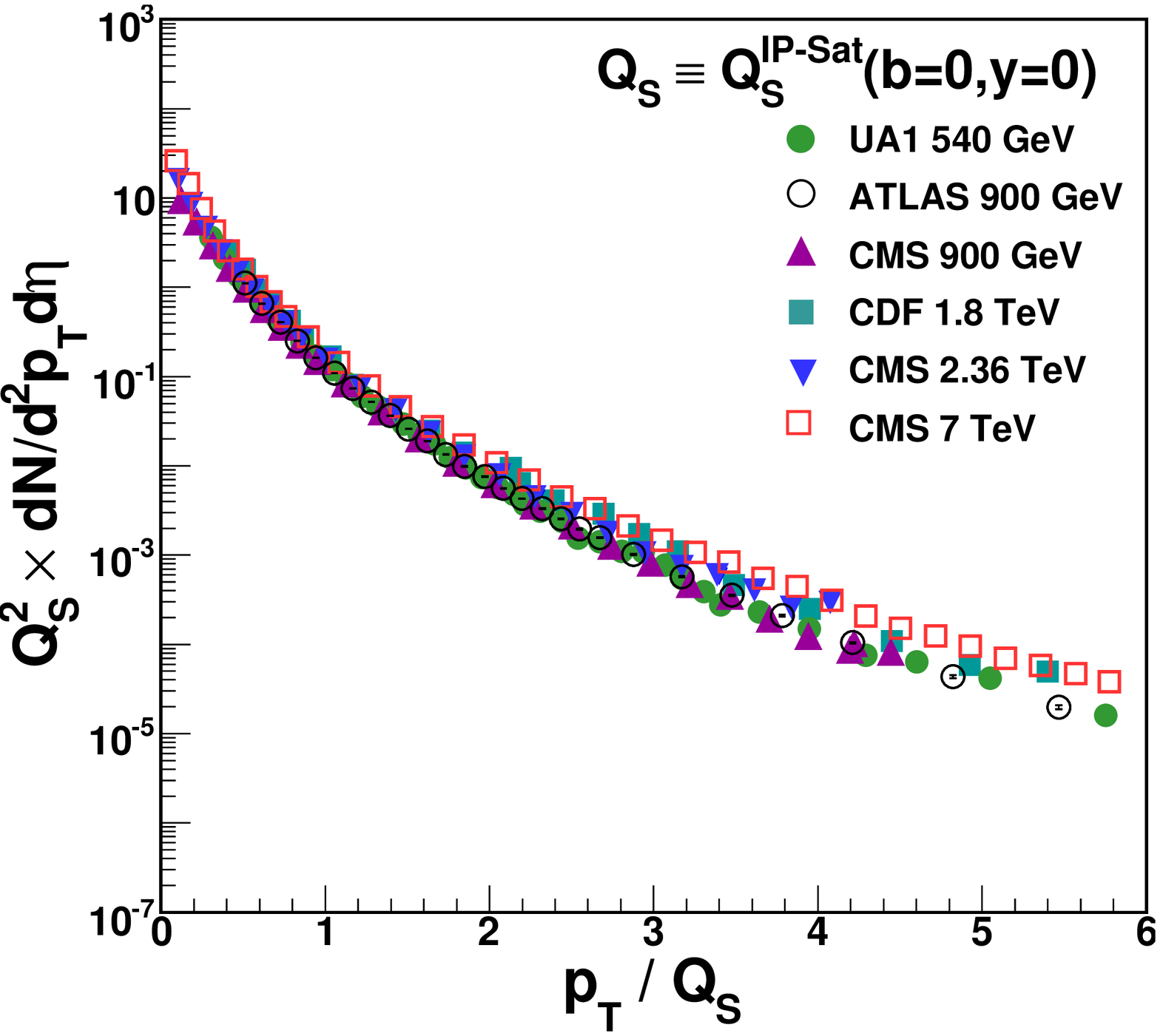}
}
\label{fig:scaling}
\caption{Left: Scaling of $\langle \pt \rangle$ with saturation scale. Right:Scaling of $\pt$-spectra plotted in terms of $\pt/Q_S$ ; data points are from Ref\cite{UA1-Arnison,CDF-Abe,CMS-Khtyn}.}
\end{figure}

Various kinematic variables exhibit scaling with the saturation scale\cite{Larry-Michal,PTRV}. The mid-rapidity multiplicity density scales with functional forms like $Q_S^2(s)$ and $Q_S^2(s)/\alpha_S(Q_S)$ whereas a linear functional form seem to provide very good fit to the energy dependence of $\langle \pt \rangle$ as shown in fig.\ref{fig:scaling}[left]. These results are  suggestive that $Q_S$ is the only scale that controls the bulk particle multiplicity. In Ref. \cite{Larry-Michal,Praszalowicz} it has been shown that $\pt$ spectra in $p+p$ collisions exhibit geometric scaling assuming a simple form of $Q_S$. In our case we use a scaling variable $\pt/Q_S$, where $Q_S$ is directly calculated in the IP-Sat model.
As shown in fig.\ref{fig:scaling}[right], an approximate scaling below $\pt/Q_S<3$ is observed for transverse momentum distribution in $p+p$ collision energy $\sqrt{s} \ge 540$ GeV. Going to lower energies we observe systematic deviations from the universal curve. \\

In summary, our description of multiplicity distribution successfully describes bulk LHC p+p data. In particular, we observe that the dominant contribution to multiplicity fluctuations is due to the  intrinsic fluctuations of gluon produced from multiple Glasma flux tubes rather than from the fluctuations in the sizes and distributions of hotspots. The $\pt$-spectra in p+p at high energies exhibits universal scaling as a function of $\pt/Q_S$. The observed scaling indicates that particle production in this regime is dominantly from saturated gluonic matter characterized by one universal scale $Q_S$. Ridge like two particle correlation structures in $\Delta\eta-\Delta \Phi$ in high multiplicity p+p collisions may provide more detailed insight into its properties~\cite{DumitDGJLV}. 

R.V was supported by the US Department of Energy under DOE Contract No.DE-AC02-98CH10886.


\begin{thebibliography}{25}
\bibitem{GLR}L.V. Gribov, E.M. Levin, M.G. Ryskin, Phys. Rept. {\bf 100}, 1 (1983); A.H. Mueller, J-W. Qiu, Nucl. Phys. {\bf B} {\bf 268}, 427 (1986).

\bibitem{MV}L.D. McLerran, R. Venugopalan,  Phys. Rev. {\bf D} {\bf 49}, 2233 (1994); {\it ibid.} {\bf 49}, 3352 (1994); {\it ibid.} {\bf 50}, 2225 (1994).

\bibitem{Iancu-RV}
  E.~Iancu and R.~Venugopalan,
  arXiv:hep-ph/0303204; F.~Gelis, E.~Iancu, J.~Jalilian-Marian and R.~Venugopalan,
  arXiv:1002.0333 [hep-ph].

\bibitem{Larry-Michal}L.~McLerran,  M.~Praszalowicz,
  Acta Phys.\ Polon.\  B {\bf 41}, 1917 (2010); {\it ibid.} {\bf 42}, 99 (2010).

\bibitem{Lev-Rez}E.~Levin, A.~H.~Rezaeian,
  Phys.\ Rev.\  D {\bf 82}, 014022 (2010).
\bibitem{PTRV}
  P.~Tribedy and R.~Venugopalan,
  Nucl.\ Phys.\  A {\bf 850}, 136 (2011)
  [arXiv:1011.1895 [hep-ph]]
\bibitem{Praszalowicz}
  M.~Praszalowicz,
  arXiv:1101.0585 [hep-ph].

\bibitem{KT}J.~Bartels, K.~J.~Golec-Biernat,  H.~Kowalski,
  Phys.\ Rev.\  D {\bf 66}, 014001 (2002); H. Kowalski, D. Teaney, Phys. Rev. D 68, 114005
(2003).

\bibitem{KMW}H.~Kowalski, L.~Motyka, G.~Watt,
  Phys.\ Rev.\  D {\bf 74}, 074016 (2006).

\bibitem{IIM}E.~Iancu, K.~Itakura, S.~Munier,
  Phys.\ Lett.\  B {\bf 590}, 199 (2004).

\bibitem{KW}G.~Watt, H.~Kowalski,
  Phys.\ Rev.\  D {\bf 78}, 014016 (2008)

\bibitem{DumitGMV}A. Dumitru, F. Gelis, L. McLerran, R. Venugopalan, Nucl. Phys. {\bf A} {\bf
  810}, 91 (2008).

\bibitem{GelisLM}F. Gelis, T. Lappi,  L. McLerran, Nucl. Phys. A828 (2009) 149.
\bibitem{LappiSV}{T. Lappi, S. Srednyak, R. Venugopalan}, JHEP {\bf 1001} 066 (2010).

\bibitem{UA1-Arnison}
  G.~Arnison {\it et al.}  [UA1 Collaboration],
  Phys.\ Lett.\  B {\bf 118}, 167 (1982).
 
\bibitem{CDF-Abe}
  F.~Abe {\it et al.}  [CDF Collaboration],
  Phys.\ Rev.\ Lett.\  {\bf 61}, 1819 (1988).


\bibitem{CMS-Khtyn}
  V.~Khachatryan {\it et al.}  [CMS Collaboration],
  Phys.\ Rev.\ Lett.\  {\bf 105}, 022002 (2010).

\bibitem{ALICE-Aamodt}
  K.~Aamodt {\it et al.}  [ALICE Collaboration],
  Eur.\ Phys.\ J.\  C {\bf 68}, 345 (2010).

\bibitem{DumitDGJLV}A.~Dumitru, K.~Dusling, F.~Gelis, J.~Jalilian-Marian, T.~Lappi, R.~Venugopalan,
  arXiv:1009.5295 [hep-ph].

\end{thebibliography}
\end{document}